\documentclass[aps,twocolumn,prl,superscriptaddress,floatfix,longbibliography]{revtex4-2}

\usepackage{amssymb,amsmath,amsthm,bbm,bbold}
\usepackage{amsfonts}
\usepackage{graphicx}
\usepackage{mathtools}
\usepackage{natbib}
\usepackage[breaklinks]{hyperref}
\usepackage[normalem]{ulem}
\usepackage{color}
\usepackage{xcolor}

\newcommand{\g}{\gamma}

\newcommand{\Tr}{\mbox{Tr}}
\newcommand{\bra}[1]{\mbox{$\langle #1 |$}}
\newcommand{\ket}[1]{\mbox{$| #1 \rangle$}}
\newcommand{\F}{\mathcal{F}}
\newcommand{\bd}[1]{\mbox{$\boldsymbol{#1}$}}

\begin{document}

\author{Lexin Ding}
\affiliation{Faculty of Physics, Arnold Sommerfeld Centre for Theoretical Physics (ASC),\\Ludwig-Maximilians-Universit{\"a}t M{\"u}nchen, Theresienstr.~37, 80333 M{\"u}nchen, Germany}
\affiliation{Munich Center for Quantum Science and Technology (MCQST), Schellingstrasse 4, 80799 M{\"u}nchen, Germany}

\author{Julia Liebert}
\affiliation{Faculty of Physics, Arnold Sommerfeld Centre for Theoretical Physics (ASC),\\Ludwig-Maximilians-Universit{\"a}t M{\"u}nchen, Theresienstr.~37, 80333 M{\"u}nchen, Germany}
\affiliation{Munich Center for Quantum Science and Technology (MCQST), Schellingstrasse 4, 80799 M{\"u}nchen, Germany}

\author{Christian Schilling}
\email{c.schilling@physik.uni-muenchen.de}
\affiliation{Faculty of Physics, Arnold Sommerfeld Centre for Theoretical Physics (ASC),\\Ludwig-Maximilians-Universit{\"a}t M{\"u}nchen, Theresienstr.~37, 80333 M{\"u}nchen, Germany}
\affiliation{Munich Center for Quantum Science and Technology (MCQST), Schellingstrasse 4, 80799 M{\"u}nchen, Germany}

\title{Comment on ``Self-Consistent-Field Method for Correlated Many-Electron Systems with an
Entropic Cumulant Energy''}

\date{\today}

\begin{abstract}
In [Phys.~Rev.~Lett.~128, 013001 (2022)] a novel ground state method was proposed. It has been suggested that this $i$-DMFT would be a method within one-particle reduced density matrix functional theory (DMFT), capable of describing accurately molecules at various geometries with an  information-theoretical nature. We reassess this work and its suggestions from a conceptual and practical point of view, leading to the following conclusions: i) A method which assigns to each molecule $\mathcal{M}$ its own functional $\mathcal{F}_{\!\mathcal{M}}$ is not a functional theory (striking violation of ``universality'') ii) even for the simplest systems $i$-DMFT yields incorrect one-particle reduced density matrices and iii) the use of an information-theoretical concept to describe molecular dissociation limits was not essential. The latter insight may help to fix the deficiency of $i$-DMFT to not reproduce correctly the smaller occupation numbers and thus to not recover the important dynamic correlations.
\end{abstract}

\maketitle
One-particle reduced density matrix functional theory (DMFT) \cite{Gilbert75,Levy79,Valone80} is a method for solving the ground state problem for Hamiltonians  $H(h)\equiv h+W$, with $W$ the fixed Coulomb pair interaction: The minimization of $\Tr[h \g]+ \F(\g)$ yields the energy $E(h)$ and one-particle reduced density matrix (1RDM) $\g(h)$ of the ground state of $H(h)$, \emph{for any} one-particle Hamiltonian $h$. The functional $\F(\g)$ is \emph{universal} since it depends only on $W$ and the particle number $N$, but not on $h$.

\begin{figure}[htb]
\centering
\includegraphics[scale=0.325]{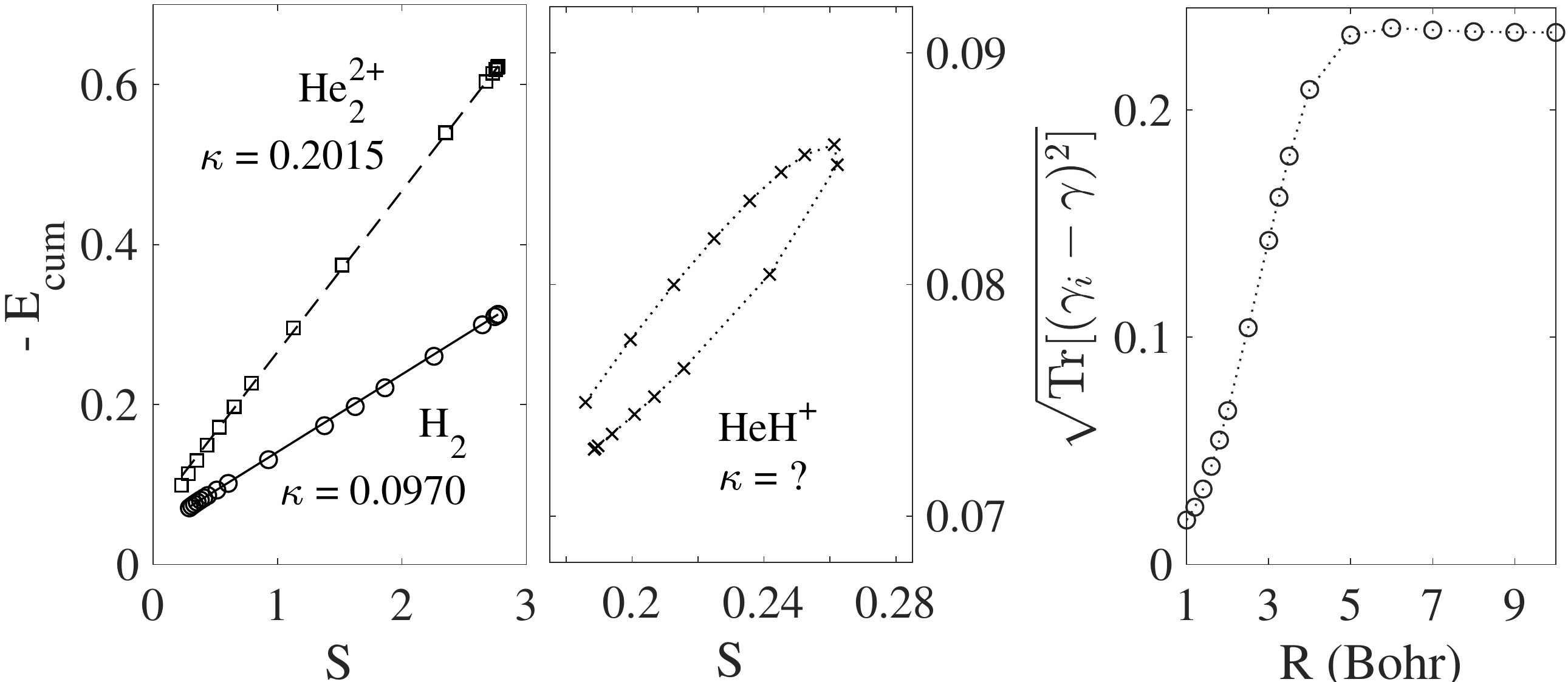}
\caption{Cumulant energy $E_\text{cum}(R)$ versus particle-hole symmetric entropy $S(\bd{n}(R))$ for the ground states of $\mathrm{H_2}, \mathrm{He_2^{2+}}$ (left) and $\mathrm{HeH^{+}}$ (middle) for different nuclear separations $R$. Right: Error of the $i$-DMFT 1RDM $\g_i$
measured by the Frobenius norm $\sqrt{\Tr[(\g_i-\g)^2]}$ with $\g$ the exact/FCI 1RDM. The same basis sets were used as in \cite{WangBaerends22-PRL} (cc-pVDZ).}\label{fig}
\end{figure}

In Ref.~\cite{WangBaerends22-PRL} a novel ground state method was proposed. It has been suggested that this $i$-DMFT would be (i) a method within DMFT (ii) capable of describing accurately molecules at various geometries, with a particular emphasis on the dissociation limit and (iii) based on a distinctive information-theoretical approximation of the 2RDM cumulant \cite{Kutz99}.
The results of \cite{WangBaerends22-PRL} are certainly remarkable, as they hinge on a possible linear dependence, $E_{\text{cum}} \approx  -\kappa \,S - b$, of the cumulant energy $E_{\text{cum}}$ and the von Neumann entropy \cite{WangKnowles21-PRA} or its particle-hole symmetric variant $S(\bd{n})= -\sum_{i} \left[n_i \ln{(n_i)}+(1-n_i) \ln{(1-n_i)}\right]$ \cite{WangBaerends22-PRL}.
Yet, the key statements (i)-(iii) are not justified and require a concise reassessment in order to not mislead the future development of DMFT:
\paragraph*{Statement (i).---} The functional approximation proposed in \cite{WangBaerends22-PRL} reads
\begin{equation}\label{func}
\F(\bd{n},\bd{\chi}) = Y(\bd{n},\bd{\chi})-\kappa \,S(\bd{n})-b\,,
\end{equation}
where $Y$ denotes the Hartree-Fock functional \cite{LiebHF,PG16} and $\g$ is parameterized by its eigenvalues $\bd{n}\equiv (n_i)$ and eigenstates $\bd{\chi}\equiv (\chi_i)$. For each of the three molecules $\mathrm{H_2, H_2O}$ and $\mathrm{N_2}$ the parameters $\kappa, b$ were then fitted such that \mbox{$i$-DMFT} yields the exact energy at the equilibrium geometry and in the dissociation limit. Arguably, the difference between the resulting parameter sets $(\kappa,b)$ does not contradict the universality of $\F$ since all three molecules have \emph{different} electron numbers. Yet, our numerical analysis of three $2$-electron molecules presented in Fig.~\ref{fig} comprehensively does: the corresponding parameter sets $(\kappa,b)$ do not coincide. Even worse, for $\mathrm{HeH^+}$ there does not exist a single-valued dependence of $E_{\text{cum}}$ on $S$ and thus $i$-DMFT would fail to describe its chemical behavior. Hence, $i$-DMFT is not a method \emph{within DMFT}. On the contrary, it assigns to each molecule $\mathcal{M}$ its own functional $\F_{\!\mathcal{M}}$. The latter still depends, however,
on $d^2$ variables ($d$ the dimension of the one-particle Hilbert space) despite the fact that it reproduces the correct energies of only a one- or two-dimensional manifold of Hamiltonians. In that sense, \mbox{$i$-DMFT} violates the common duality principle \cite{Levy79,Lieb83} which  suggests that $\F_{\mathcal{M}}$ should depend on only one or two parameters.

\paragraph*{Statement (ii).---} As it is confirmed in the right panel of Fig.~\ref{fig}, $i$-DMFT yields rather incorrect 1RDMs, even for the simple hydrogen molecule $\mathrm{H_2}$.
This is not surprising since the underlying expansion \eqref{func} refers to the regime of weak correlations \cite{Levy87}. To explain this conceptual flaw of $i$-DMFT, let us assume that it yielded for $\mathrm{H}_2$ in the dissociation limit the \emph{correct} 1RDM, i.e., $\g= \frac{1}{2}\sum_{i=L/R, \sigma =\uparrow/\downarrow}\ket{1s_i\sigma}\!\bra{1s_i\sigma}$, where $\ket{1s_{L/R}\sigma}$ denotes the 1s hydrogen orbital at the left/right nucleus. Then, the resulting Hartree-Fock term $Y$ would erroneously describe interaction effects between two electrons in the 1s hydrogen orbital \emph{at the same nucleus}.

\paragraph*{Statement (iii).---} The suggested information-theoretical nature of $i$-DMFT is not obvious:
In the dissociation limit, e.g., the sole purpose of $-\kappa S(\bd{n})$ is to enforce --- in contrast to the equally exact unrestricted Hartree-Fock ($Y$) \cite{LiebHF,Lieb94HF} --- the desired degree of mixedness of the 1RDM. Since the one-particle Hamiltonian restricts for $\mathrm{H}_2$ the active space effectively to only four spin-orbitals, other particle-hole symmetric Schur-convex functions $f(\bd{n})$ added with an appropriate weight to $Y$ in \eqref{func} would have a similar effect. This is indeed the case since Schur-convex functions attain their minimum at the vector with maximal mixedness, here $(n_1,\ldots,n_4)=1/2\,(1,1,1,1)$. Such a replacement of $n_i \ln{(n_i)}$ in $S(\bd{n})$ could be chosen to regain the correct scaling $\partial\F/\partial n_i \sim -1/\sqrt{n_i}$ for $n_i \approx 0$ (``exchange force'') \cite{LS56,Cs17stability,CS19ExForce}. This would also fix the deficiency of $i$-DMFT to not reproduce correctly the smaller $n_i$ (c.f Table VII in the SI of \cite{WangBaerends22-PRL}), their decaying behaviour and in that sense the important dynamic correlations.

\begin{acknowledgments}
We thank P.Knowles for instructive discussions. We also acknowledge financial support from the German Research Foundation (Grant SCHI 1476/1-1) (L.D., J.L., C.S.), the Munich Center for Quantum Science and Technology (L.D., C.S.) and the International Max Planck Research School for Quantum Science and Technology (IMPRS-QST) (J.L.). The project/research is part of the Munich Quantum Valley, which is supported by the Bavarian state government with funds from the Hightech Agenda Bayern Plus.
\end{acknowledgments}

\bibliography{refs}

\end{document}